\documentclass[a4paper, 11pt]{article}
\usepackage[affil-it]{authblk} 
\usepackage{etoolbox}
\usepackage{lmodern}
\usepackage[a4paper]{geometry}
\usepackage[in]{fullpage}
\usepackage[utf8]{inputenc}
\usepackage{amsmath}
\usepackage{amssymb}
\usepackage{cite}

\usepackage{graphicx}

\numberwithin{equation}{section}

\usepackage{footnote}
\usepackage{amsfonts}
\usepackage{graphicx}
\usepackage[dvipsnames]{xcolor}
\usepackage[colorlinks]{hyperref}
\hypersetup{linkcolor=blue,citecolor=blue,urlcolor=blue}

\usepackage{subfiles}

\DeclareMathAlphabet{\mathpzc}{OT1}{pzc}{m}{it}
\DeclareMathAlphabet{\mathcalligra}{T1}{calligra}{m}{n}

\usepackage[framemethod=TikZ]{mdframed}
\mdfdefinestyle{MyFrame2}{%
	linecolor=blue!20,
	middlelinewidth=2pt,
	frametitlerule=true,
	frametitlebackgroundcolor=blue!20,
	frametitlerulecolor=blue!20,
	frametitlerulewidth=1pt,
	innertopmargin=\topskip}
	
\title{Superfluid dark matter flow around cosmic strings}

\makeatletter
\patchcmd{\@maketitle}{\LARGE \@title}{\fontsize{16}{19.2}\selectfont\@title}{}{}
\makeatother

\author[1]{Heliudson Bernardo\footnote{Email: \href{mailto:heliudson@hep.physics.mcgill.ca}{heliudson@hep.physics.mcgill.ca}}}

\author[1,2]{Robert Brandenberger\footnote{Email: \href{mailto:rhb@physics.mcgill.ca}{rhb@physics.mcgill.ca}}}

\author[1]{Aline Favero\footnote{Email: \href{mailto:aline.favero@mail.mcgill.ca}{aline.favero@mail.mcgill.ca}}}

\affil[1]{Department of Physics, McGill University, Montreal, QC, H3A 2T8, Canada}

\affil[2]{Institute for Theoretical Physics, ETH Z\"urich, CH-8093 Z\"urich, Switzerland}

\date{\vspace{-5ex}}

\begin{document}

\maketitle

\begin{abstract}
    We consider a cosmic string moving through a gas of superfluid dark matter (SFDM) particles and analyze how it affects the dark matter distribution. We look at two different cases: first, a cosmic string passing through an already condensed region, and second, through a region that is not yet condensed. In the former, the string induces a weak shock in the superfluid, and the Bose-Einstein condensate (BEC) survives. In the latter, a wake of larger density is formed behind the string, and we study under which conditions a BEC can be formed in the virialized region of the wake. By requiring the thermalization of the DM particles and the overlap of their de Broglie wavelengths inside the wake, we obtain an upper bound on the mass of the dark matter particles on the order of 10 eV, which is compatible with typical SFDM models.
\end{abstract}

\section{Introduction}

The fundamental nature of dark matter (DM) remains a significant open problem in cosmology. Although the $\Lambda$CDM model successfully describes it, on large scales, as a fluid of collisionless particles and vanishing sound speed, small DM interactions with itself or other species would affect its small-scale distribution and behavior. In fact,
there is a striking correlation between the acceleration in galaxies and the total baryonic mass within them \cite{Lelli:2016cui, DiPaolo:2018mae}. These and other correlations \cite{Lelli:2016uea} challenge the cold DM picture on galactic scales \cite{Bullock:2017xww} but can be accounted for by the superfluid dark matter approach \cite{Khoury:2016ehj}.

Superfluid dark matter models are essentially based on the formation of a Bose-Einstein condensate (BEC) of DM particles on galactic scales and its associated phonon excitations \cite{Sin:1992bg,Goodman:2000tg, Peebles:2000yy, Hu:2000ke, Silverman:2002qx, Boehmer:2007um, Chavanis:2011zi,Bettoni:2013zma, Guth:2014hsa, Berezhiani:2015bqa, Berezhiani:2015pia,Hui:2016ltb, Ferreira:2018wup} (see also \cite{Das:2014agf,Das:2018udn, Das:2022mgr} for studies of cosmological BEC and \cite{Ferreira:2020fam, Hui:2021tkt, Khoury:2021tvy} for a more complete list of references). The phonons might mediate a long-range interaction between the baryons, which reproduce the modified Newtonian dynamics (MOND) of the baryonic acceleration on such scales \cite{Berezhiani:2017tth}. Given their typical speeds in galaxies, the (supposedly bosonic) DM particles should have a mass of the order of eV or less for the BEC to be formed \cite{Berezhiani:2015pia}. Moreover, thermal effects should be included so that the condensed region is a genuine superfluid with inviscid and normal flows, and the density profile of finite temperature superfluid cores within galaxies can be computed for different superfluid equations of state \cite{Sharma:2018ydn}.

Finite temperature superfluids are described by the two-fluid model historically initiated by London and Tisza \cite{London,london1938lambda, tisza1938transport} but independently established by Landau \cite{Landau} (see also \cite{Donnelly, BALIBAR2017586} for historical notes). In this model, a superfluid has two associated flows, an inviscid and a normal one, in which entropy and temperature can only be transported by the latter. One commonly talks about a ``two-component'' fluid where the superfluid component is a BEC, and the normal component is composed of quasiparticles in thermal equilibrium \cite{LLfluids, Wilks_1957}. The relative energy density in these components depends on the temperature, and only the normal component is present for temperatures greater than the BEC critical temperature. The Landau two-fluid model predicted two kinds of sounds in superfluids: the usual adiabatic one, where perturbations in the energy density and pressure are propagated with sound speed $c_1$, and another one associated with the propagation of entropy and temperature fluctuations with sound speed $c_2$. Explaining these sound speeds for Helium II is one of the great achievements of the Landau two-fluid model.

Superfluid DM would also exhibit wave-like interference patterns on large scales \cite{Schive:2014dra,Mocz:2019pyf} and has also been proposed as an explanation for the origin of cosmic filament spin \cite{Alexander:2021zhx}. The fundamental differences in the physics of cold and superfluid DM translate to new prospects for observations. In the present work, we take a step toward understanding how superfluid DM modifies cosmic string signatures \cite{Kibble:1976sj, Vilenkin:1984ib, Brandenberger:1993by, vilenkin1994cosmic, Hindmarsh:1994re}.

It is well-known how the motion of a long cosmic string through a gas of DM particles affects its distribution. For collisionless DM, a wake is formed downstream of the flow, with the cosmic string at its apex, within which the density is twice the initial DM density \cite{Silk:1984xk,Rees,Vachaspati:1986zz,Stebbins:1987cy}. These cosmic string wakes affect the accretion of baryonic matter in a statistically different way compared to the usual DM accretion, leading to distinct signatures \cite{Kaiser:1984iv,Moessner:1993za,Perivolaropoulos:1994ry,Magueijo:1995xj,Pen:1997ae,Danos:2010gx,Brandenberger:2010hn,Hernandez:2012qs,NevesdaCunha:2018urr,Laliberte:2018ina,Maibach:2021qqf,Blamart:2022kly}. In this paper, we investigate how this picture changes if the DM particles are light bosons that can undergo a phase transition and generate a BEC\footnote{Natural ultralight DM candidates are axion fields, which can also have cosmic string solutions. For generality, we shall not assume a relation between the field theory origin of DM and the cosmic string. For a study on cosmic strings with BEC in their core, see \cite{Harko:2014pba}.}. 

We study two situations where the superfluid phase of DM particles can generate differences in the flow around moving cosmic strings compared to cold DM: firstly, the moving cosmic string can pass by an already condensed region; and secondly, the wake formation can happen at redshifts such that the DM density inside the wake is greater than a critical density necessary for a BEC. In the latter case, the DM particles might condense inside the virialized region in the wake; in the former, the cosmic string wake will generically contain a shock since cosmic string speeds are relativistic and, as such, supersonic relative to the superfluid.

After some reasonable physical assumptions, this paper provides first-principle analytic computations of superfluid DM flows around moving cosmic strings. For generality, and since cosmic strings move at relativistic speeds, we consider the relativistic effective field theory approach to superfluids\footnote{See \cite{Berezhiani:2019pzd,Berezhiani:2020umi} for limitations of the non-relativistic effective field theory approach in dealing with supersonic processes.}. Relevant aspects of this description are reviewed in the next section. In section \ref{shocks}, we solve the Taub-Rankine-Hugoniot junction equations for linearized and strong shocks in the superfluid. In section \ref{superfluidsinwakes}, we estimate the redshift where BEC can be formed in the cosmic string wake as a function of the DM particle mass. We discuss applications and prospects for future directions in section \ref{conclusion}.

\section{Field theory approach to superfluids} \label{EFT_superfluids}

There are different but equivalent formalisms generalizing Landau's two-fluid model to the relativistic case \cite{ISRAEL1981, KHALATNIKOV1982,lebedev1982relativistic,ISRAEL1982_2,Carter:1992gmy,Carter:1995if,Son:2000ht,Nicolis:2011cs} (see also \cite{Herzog:2008he, Alford:2012vn} and references therein). To set some notation and make the discussion self-contained, in this section, we review some of the results on the effective field theory (EFT) description of superfluids as presented in \cite{Nicolis:2011cs}, which we will follow closely. For simplification, we will assume that the normal component is dissipationless.

From the EFT perspective, below the superfluid's critical temperature, $T_c$, the superfluid component is a BEC described by a state that spontaneously breaks a global U(1) symmetry. This U(1) symmetry is associated with the conservation of particle number in the fluid. By the Goldstone theorem, there is a gapless excitation $\psi$ which non-linearly realizes the U(1) symmetry as
\begin{equation}
    \psi \to \psi +a, \quad a = \text{const.}
\end{equation}
The most general action that is Poincar\'e invariant and compatible with this symmetry has the form
\begin{equation}
    S = \int d^4x F(X), \quad \mathrm{with} \;X = \partial_\mu \psi \partial^\mu \psi,
\end{equation}
and the U(1) current is given by 
\begin{equation}
    j^\mu (x) = 2F'(X)\partial^\mu \psi.
\end{equation}

A homogeneous and isotropic BEC state (or superfluid phase at $T=0$) can be described by $\psi= \mu t$, since this implies a state of uniform charge density and vanishing spatial current,
\begin{equation}
    j^\mu = -2\mu F'(-\mu^2) \left(1,0,0,0\right).
\end{equation}
The equation of motion for $\psi$ is equivalent to the conservation of particle number, $\partial_\mu j^\mu = 0$.  Note that if we define a four-velocity satisfying $u^\mu \propto \partial^\mu \psi$, the system is equivalent to a fluid with an irrotational flow. Now, at finite temperature, there will be excitations in the fluid/gas, and not all particles will be condensed in the ground state: some particles will, instead, occupy excited states. At finite $T$, these perturbations will reach thermal equilibrium, and if they are in a regime where the mean free time and path of a phonon are much smaller than the spacetime volume occupied by the fluid, this thermal bath of phonons will be described by usual hydrodynamics. However, as $T$ approaches $T_c$, the symmetry is restored, the BEC is gone and only the normal component remains.

To describe the normal fluid component in field theory, we use embedding coordinates $\phi^i$ of the fluid. They map spacetime  points to positions of the fluid elements, $x^\mu \to \phi^i(x,t)$, $i=1,2, 3$ \cite{Dubovsky:2005xd, Dubovsky:2011sj, Andersson:2020phh}. At a fixed time, these maps should be invertible ($\det \partial_i \phi^i\neq 0$), and, if the fluid is incompressible, also volume preserving ($\det \partial_j \phi^i =1$). For dissipationless fluids, given Poincar\'e invariance and the homogeneity and isotropy of the fluid's internal space, the $\phi^i$ should enter the action through the combination \cite{Dubovsky:2005xd}
\begin{equation}
    J^\mu = \frac{1}{6}\epsilon^{\mu\alpha\beta\gamma}\epsilon_{ijk}\partial_\alpha \phi^i \partial_\beta \phi^j \partial_\gamma \phi^k.
\end{equation}
With $\partial^\mu \psi$ and $J^\mu$, we can construct three scalar quantities \cite{Nicolis:2011cs}, 
\begin{equation}
    X = \partial_\mu \psi \partial^\mu \psi, \quad b = \sqrt{-J^\mu J_\mu}, \quad \text{and}\; y = -\frac{1}{b} J^\mu \partial_\mu \psi.
\end{equation}
The vector $u^\mu =- J^\mu/b$ is actually the four-velocity of the normal fluid component, since it is normalized to $-1$ and the fluid's comoving coordinates do not change along its integral curves, $u^\mu \partial_\mu \phi^i = 0$.

In summary, the low-energy Lagrangian density describing superfluidity will have the form
\begin{equation}\label{Lagdensity}
    \mathcal{L} = F(b, X, y).
\end{equation}
All the infrared dynamics of finite-temperature relativistic superfluids are encoded in that Lagrangian. It can be shown that one can recover two sound speeds for perturbations around an equilibrium solution for $\psi$ and $\phi^i$, which are the relativistic versions of Landau's two sounds  \cite{Nicolis:2011cs}. 

To recover the superfluid's hydrodynamics, we first write the superfluid action for an arbitrary metric
\begin{equation}
    S = \int d^4x \sqrt{-g} \:F(b,X,y)
\end{equation}
and then compute its associated energy-momentum tensor
\begin{equation}\label{EMtensor}
    T_{\mu\nu} = 2XF_X \tilde{u}_\mu \tilde{u}_\nu + (yF_y -bF_b)u_\mu u_\nu + (F-bF_b)g_{\mu\nu},
\end{equation}
where $\tilde{u}_\mu = -\partial_\mu \psi /\sqrt{-X}$. Note that $X$, $b$, and $y$ now depend on the metric.

Moreover, the current associated with the U$(1)$ symmetry $\psi \to \psi + \text{const.}$ is
\begin{equation}\label{Ncurrent}
    j^\mu = -2\sqrt{-X}F_X\tilde{u}^\mu + F_y u^\mu,
\end{equation}
and its conservation physically means the conservation of particle number. Both $T_{\mu \nu}$ and $j^\mu$ have contributions from the superfluid and normal components.

Now, we can start identifying the fluid variables. In the frame comoving with the normal fluid, we have
\begin{equation}
    T_{\mu\nu} u^\mu u^\nu = yF_y -F -2y^2F_X = \rho
\end{equation}
and
\begin{equation}
    -j^\mu u_\mu = F_y -2yF_X = n,
\end{equation}
where, by definition, $\rho$ and $n$ are the energy and number densities that can be measured in that frame. To identify the fluid stresses, we contract the energy-momentum tensor with the projector associated to the spacelike directions perpendicular to $u^\mu$ ,
\begin{equation}
    T^{\mu\nu}(\eta_{\mu\nu} +u_\mu u_\nu) = 3(F-bF_b) - 2F_X (X+y^2).
\end{equation}
So, we identify $p = F- bF_b$ as the pressure, and the term $\propto (X+y^2)$ as a contribution from the anisotropic stress-tensor (this could also be seen directly from $T_{\mu\nu}$). The chemical potential $\mu$, entropy density $s$ and temperature $T$ are identified after assuming the first law of thermodynamics \cite{Nicolis:2011cs} (see also \cite{Ballesteros:2012kv}):
\begin{equation}
    \mu = y, \quad s = b, \quad \mathrm{and} \quad T= -F_b.
\end{equation}
The entropy current
\begin{equation}
    J^\mu  = s u^\mu  = b u^\mu,
\end{equation}
is identically conserved (as expected for a non-dissipative fluid).

Computing the differential of $p$, we obtain \cite{Nicolis:2011cs}
\begin{equation}
    dp = s dT + n d\mu + 2F_X \xi d\xi,
\end{equation}
where $\xi$ is the modulus of the spacelike four-vector $\xi^\mu = (\eta^{\mu\nu} +u^\mu u^\nu)\partial_\nu \psi$.
So, the pressure is a function of $T$, $\mu$, and $\xi$. This defines the equation of state of the superfluid. Once this function is given, we can construct the Lagrangian as
\begin{equation}
    \mathcal{L} = F = p + b F_b = p- sT = p- T \frac{dp}{dT},
\end{equation}
while expressing the result as a function of $b = s = dp/dT$, $y = \mu$, and $X = \xi^2 -y^2$. This maps the thermodynamic and field theory descriptions, and vice-versa.

To better understand the physical meaning of $\xi^\mu$, recall first that $\tilde{u}^\mu = - \partial^\mu \psi/ \sqrt{-X}$, and so
\begin{equation}
    \tilde{u}^\mu = \frac{y}{\sqrt{-X}}u^\mu - \frac{1}{\sqrt{-X}} \xi^\mu .
\end{equation}
Contracting with $u^\mu$ gives
\begin{equation}
    \gamma = - \tilde{u}^\mu u_\mu  =  \frac{y}{\sqrt{-X}},
\end{equation}
where $\gamma$ is the Lorentz factor for the velocity of one component as measured in the other component's frame. On the other hand, contracting with $\tilde{u}^\mu$ gives
\begin{equation}
    -1 = \tilde{u}^\mu \tilde{u}_\mu = -\gamma^2 - \frac{1}{\sqrt{-X}} \xi^\mu \tilde{u}_\mu.
\end{equation}
In the normal component frame, $u^\mu = (1,0)$, $\xi^\mu = (0, \xi^i)$, and $\tilde{u}^\mu = \gamma(1, v^i)$, where $v^i$ is the superfluid component velocity. Thus,
\begin{equation}
    \gamma^2 -1 = -\frac{1}{\sqrt{-X}}\gamma \:\xi^i v_i\implies  \gamma^2 v^2 = -\frac{\gamma}{\sqrt{-X}} \xi^i v^i,
\end{equation}
and we conclude that $\xi^i = -y v^i$. When there is no relative velocity between the normal and superfluid parts, $\xi^\mu = 0$ and $y = \sqrt{-X}$.

\section{Shocks in the superfluid flow}
\label{shocks}

An interesting property of having a fluid described by two velocity fields is that it will generically have anisotropies. This can be seen, for instance, from the spatial components of $T_{\mu\nu}$ above: as long as $u^\mu \neq \tilde{u}^\mu$ there will be anisotropic stresses. Moreover, there is a non-vanishing momentum flux density in one of the frames comoving with one of the fields. For instance, in the frame where $u^\mu = (1,0,0,0)$, we have
\begin{equation}
    T^{i0} = 2X F_X \gamma^2 v^i,
\end{equation}
where $v^i$ is the relative velocity between the superfluid and normal components as measured in the latter's frame, and $\gamma$ is its associated Lorentz factor. 

In the following, we shall redefine the two timelike vector fields $\tilde{u}^\mu$ and $u^\mu$ in order to make the intrinsic superfluid anisotropy manifest.  This will also select the frame in which there is no momentum flux, and so in some sense, it might be thought of as the ``center of mass'' frame for the superfluid. A similar approach was considered in \cite{Letelier, bayin1986anisotropic}, but in the context of a non-interacting two-perfect fluid system, which is physically distinct from a superfluid. We define four-vectors $V^\mu$ and $W^\mu$ as
\begin{subequations}
\begin{align}
    V^\mu &= \cos \alpha \;\tilde{u}^\mu + R \sin \alpha\; u^\mu, \\
    W^\mu &= -\frac{1}{R} \sin \alpha \;\tilde{u}^\mu + \cos \alpha \;u^\mu,
\end{align}
\end{subequations}
where 
\begin{equation}
    R = \sqrt{\frac{yF_y - bF_b}{2X F_X}}.
\end{equation}
These definitions are such that 
\begin{equation}
    T^{\mu\nu} = 2X F_X V^\mu V^\nu + (yF_y -bF_b)W^\mu W^\nu + (F-bF_b)g^{\mu\nu},
\end{equation}
for any choice of $\alpha$. We shall fix $\alpha$ by demanding that $V^\mu W_\mu = 0$, 
\begin{equation}
    V^\mu W_\mu = 0 \implies \tan 2\alpha = \frac{2R}{R^2-1} u^\mu \tilde{u}_\mu = \frac{2R}{1-R^2}\gamma.
\end{equation}
In this way, $V^\mu$ is timelike while $W^\mu$ is spacelike\footnote{Actually, $V^\mu W_\mu=0$ alone doesn't fix which vector is timelike. We also need to assume $R<1$. If this is not so, one needs to redefine $R\to 1/R$, otherwise $W^\mu$ will be timelike. We have assumed, without loss of generality, that $2XF_X > yF_y-bF_b$, which is most reasonable at very low-temperatures.}, and there is no momentum flux in the frame where $V^i =0$ because in this case, we should have $W^0 = 0$, and so $T_{i0} = 0$. Note that the Lorentz factor after the last equality is associated with the relative velocity between $\tilde{u}^\mu$ and $u^\mu$. 

In terms of the normalized vectors
\begin{equation}
    U^\mu = \frac{V^\mu}{\sqrt{-V^\rho V_\rho}} \;\: \mathrm{and}\;\: A^\mu = \frac{W^\mu}{\sqrt{W^\rho W_\rho}}, 
\end{equation}
we have
\begin{equation}
    T_{\mu\nu} = 2X F_X(-V^\rho V_\rho)U_\mu U_\nu+(yF_y - b F_b)(W^\rho W_\rho) A_\mu A_\nu + (F-bF_b)g_{\mu\nu}.
\end{equation}
Computing $V^2$ and $W^2$, we find
\begin{equation}
    V^\mu V_\mu = -\frac{1}{2}(1+R^2) - \frac{1}{2}\frac{(1-R^2)}{\cos 2\alpha}, \quad W^\mu W_\mu = -\frac{1}{2}\left(1+ \frac{1}{R^2}\right)-\frac{1}{2}\left(1-\frac{1}{R^2}\right)\frac{1}{\cos 2\alpha},
\end{equation}
and the $(\cos 2\alpha)^{-1}$ in these expression is given by
\begin{equation}
    \frac{1}{\cos2\alpha} = \frac{\left[(2XF_X + yF_y -bF_b)^2 +4(2XF_X)(yF_y - bF_b)\left((\tilde{u}^\mu u_\mu)^2-1\right)\right]^{1/2}}{|2X F_X - (yF_y - bF_b)|}.
\end{equation}
So, we can write
\begin{equation}
    T_{\mu\nu} =(\rho_U+p_U) U_{\mu} U_\nu + p_U g_{\mu\nu} + (p_A - p_U) A_\mu A_\nu,
\end{equation}
where we have defined
\begin{align}
    \rho_U &= -F +\frac{1}{2}(2XF_X + yF_y +bF_b)+\nonumber\\
    & +\frac{1}{2}  \left[(2XF_X + yF_y -bF_b)^2 +4(2XF_X)(yF_y - bF_b)\left((\tilde{u}^\mu u_\mu)^2-1\right)\right]^{1/2}, \\
    p_A &= F -\frac{1}{2}(2XF_X +yF_y +bF_b) + \nonumber\\
    &+ \frac{1}{2} \left[(2XF_X + yF_y -bF_b)^2 +4(2XF_X)(yF_y - bF_b)\left((\tilde{u}^\mu u_\mu)^2-1\right)\right]^{1/2}, \\
    p_U &= F- bF_b.
\end{align}

Note that, in the absence of relative velocity, $\tilde{u}^\mu u_\mu = -1$, we have $p_A= p_U$, and the anisotropic term vanishes. Also in this case, $R= \tan \alpha$ and 
\begin{equation}
    V^\mu V_\mu\bigg|_{\gamma=1} = -\frac{1}{\cos^2 \alpha} = -(1+R^2), \quad W^\mu\bigg|_{\gamma=1} = 0.
\end{equation}
Moreover, the energy density $\rho_U$ and pressure $p_U$ coincide with the ones discussed in section \ref{EFT_superfluids} after setting $y = \sqrt{-X}$.

Using the same definitions in the expression for the number current, we have
\begin{align}
    j^\mu &= \sqrt{-V^2}\left(-2\sqrt{-X}F_X \cos \alpha+ \frac{F_y}{R}\sin \alpha \right) U^\mu + \sqrt{W^2}\left(2R\sqrt{-X}F_X \sin \alpha + F_y \cos \alpha\right) A^\mu\nonumber\\
    &= n_U U^\mu + j_A A^\mu,
\end{align}
where
\begin{equation}
    n_U = - j_\mu U^\mu \quad \mathrm{and} \quad j_A = j_\mu A^\mu
\end{equation}
are the particle number density and flux in the frame comoving with $U^\mu$. We see that, although there is no momentum flux in such a frame, there is still a flux of particles, which vanishes if there is no relative velocity between the components.

In this section, we study how a straight cosmic string extended along the $z$-axis and moving with constant velocity $-v_+ \partial_x$ affects a cylindrical symmetric superfluid configuration, with symmetry axis along the string. Equivalently, we shall analyze the flow in the cosmic string rest frame, where the superfluid as a whole is moving with speed $v_+$ in the $x$-axis. In such symmetric configurations, chosen to simplify the analysis, the only anisotropy is parallel to the string, and we take $A^\mu = (0,0,0,1)$ in the string comoving frame. Such symmetric configurations also include the homogeneous case 
\begin{equation}
    \psi(x) = y_0 t, \quad \phi^i(x) = b_0^{1/3}x^i\implies y(x) = y_0 =\sqrt{-X}, \quad b(x) = b_0, \quad \tilde{u}^\mu = (1,0,0,0) = u^\mu,
\end{equation}
where there is no relative velocity between the components, and there are no anisotropies at all. The resulting flow solution is also a good approximation for the one when the relative velocity is very small. Note that, in the non-relativistic regime, the relative speed between the superfluid and normal velocity fields is much smaller than the string speed.   

In the cosmic string rest frame, the metric is 
\begin{equation}
    ds^2 = -dt^2 + dr^2 + dz^2 + (1- 4 G \mu)^2 r^2 d\phi^2,
\end{equation}
where $\mu$ is the string tension.

Firstly, we shall use coordinates where the cosmic string metric is still Minkowskian, but the new axial angle $\tilde{\phi} = (1-4 G \mu) \phi$ has a deficit proportional to the string tension:
\begin{equation}
    x = r \cos[(1-4 G \mu)\phi + 4 \pi G \mu], \quad y = r\sin[(1-4G \mu)\phi+ 4\pi G \mu], \quad 0\leq \tilde{\phi}<(1-4G \mu)2\pi.
\end{equation}
In these coordinates, we have
\begin{equation}
    ds^2 = -dt^2 + dx^2 + dy^2 + dz^2,
\end{equation}
and the wedge 
\begin{equation}
    -\epsilon x\leq y \leq  \epsilon x, \quad \epsilon = \tan(4\pi G \mu),
\end{equation}
is left uncovered. The line segments $y_{\pm} = \pm \epsilon x$, for $x>0$, correspond to $\phi =0$ and $\phi =2\pi$, and so should be identified. In other words, a total wedge angle of $8\pi G \mu$ is missing in the conical spacetime transverse to a cosmic string. Note that the shift $4\pi G \mu$ inside the argument of the trigonometric functions in the definition of the Cartesian coordinates only sets the position of the wedge, which in this case is along the positive $x$ axis.

When the string passes by a gas of collisionless DM particles with a spatial axial distribution along the string, individual particles in the $y>0$ ($y<0$) plane receive an impulse towards the negative (positive) $y$ axis. This produces a wake with a total aperture angle $8\pi G \mu \gamma_+$ (for small $G \mu$ and order-one Lorentz gamma factors), within which the mass density is initially twice the density outside the wake \cite{Stebbins:1987cy}. However, for a fluid with finite sound speed $c_s$, the flow around the cosmic string is more involved and strongly depends on the ratio $c_s/v$ between the sound and string speeds. In fact, for supersonic string speeds, the flow exhibits a shock, i.e. a surface of discontinuity in the fluid variables downstream of the flow past the string. This is similar to the shocks in the flow of baryonic matter, as studied in \cite{Stebbins:1987cy}, where shells of baryonic matter collide due to the relative velocity induced by the string. Note, however, that such shocks differ from the later ones that originate from gravitational instabilities inside the wake \cite{Sornborger_1997}. For superfluids at finite temperatures, there are always two intrinsic sound speeds, but we shall see that only the adiabatic sound speed is relevant for the presence of shocks. 

Since cosmic string speeds are relativistic, we shall use the formalism of relativistic shocks to describe the jump in energy density, velocity, and pressure across the shock \cite{Taub1948,Saini1961, Lichnerowicz1966, Lichnerowicz1967, Lichnerowicz_1970, Thorne1973} (see also \cite{LLfluids}). Cosmic string induced shocks were first studied in \cite{Stebbins:1987cy}, although in the non-relativistic setting. A relativistic analysis for perfect fluids with a polytropic equation of state was carried out in \cite{Deruelle:1988jg}, for the strong shock case. In the following, we shall adopt the more general results of \cite{Hiscock:1988zn}, where shocks in relativistic perfect fluids were studied and, for linearized shocks, no equations of state were assumed.

The fundamental equations for relativistic shocks are obtained from integrating the local conservation equations
\begin{equation}
    \nabla^\mu T_{\mu\nu} = 0, \quad \nabla_\mu j^\mu =0
\end{equation}
across the shock front. This results into
\begin{subequations}
\begin{align}
    \left[j^\mu \right] \mathcalligra{n}_\mu&=0, \\
    \left[T^{\mu\nu}\right] \mathcalligra{n}_{\nu} &=0,
\end{align}
\end{subequations}
where $[A] = A_+ - A_-$ denotes the difference between the value of a variable $A$ in front of and behind the shock, and $\mathcalligra{n}^\mu$ is the unit vector normal to the discontinuity surface. These are the relativistic (and covariant) versions of the Rankine-Hugoniot junction condition equations \cite{LLfluids}.

\begin{figure}[t]
	\centering
	\includegraphics[scale=.6]{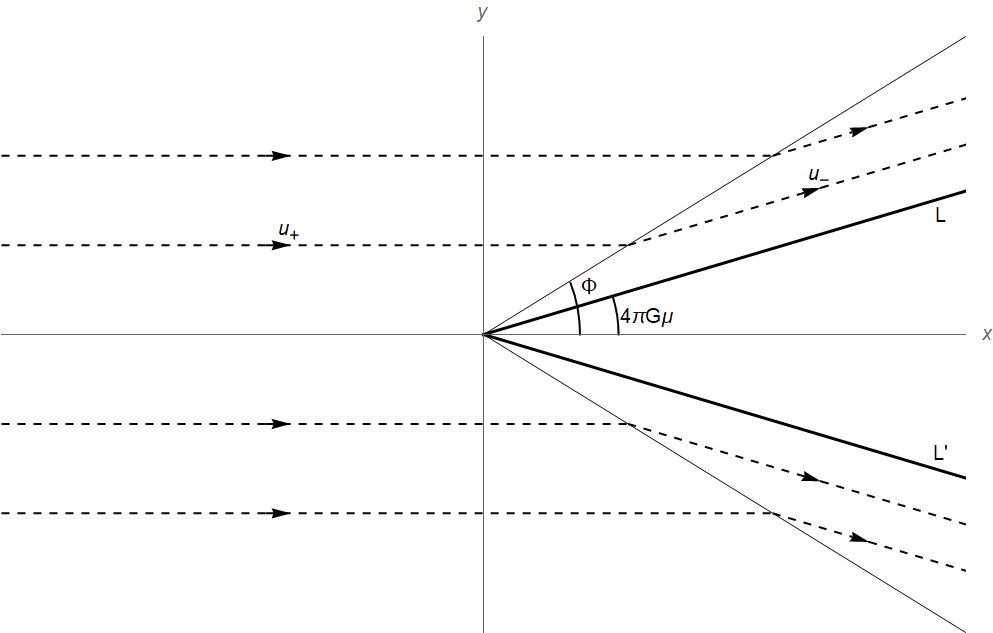}
	\caption{This figure shows a shock formed behind a cosmic string: the string is lying along the $z$-axis and $\Phi$ is the shock's half-angle. The lines $L$ and $L^{\prime}$ are equivalent, and $4 \pi G \mu$ is the missing wedge's half-angle. The dashed lines represent the fluid's streamlines.}
	\label{f01}
\end{figure}

The stationary physical configuration considered is depicted in Figure~\ref{f01}. For the cosmic string solution above, the shock front is perpendicular to the $xy$-plane and has a normal vector of the form
\begin{equation}
    \mathcalligra{n}^\mu = (0, -\sin \Phi, \cos \Phi, 0),
\end{equation}
where the half-angle of the shock $\Phi$ is to be determined. For the symmetric superfluid configurations discussed before we have $X^\mu \mathcalligra{n}_\mu = 0$, and the Taub-Rankine-Hugoniot equations give
\begin{subequations}
\begin{align}\label{RHCSeq1}
    \left[(\rho_U+p_U) U^{\mu} U^\nu + p_U \eta^{\mu\nu}\right] \mathcalligra{n}_\nu &= 0,\\
    \left[n_U U^\mu\right] \mathcalligra{n}_\mu&=0. \label{RHCSeq2}
\end{align}
\end{subequations}
We shall drop the sub-indices $U$ in the fluid variables for the rest of this section.

In the stationary case, the velocity field $U^i$ (in Cartesian coordinates) is parallel to the $x$-axis before the shock front and to the wedge after the shock front,
\begin{equation}
    U^\mu_+ = (\gamma_+, U_+, 0, 0), \quad U^\mu_- = (\gamma_-, U_-, \epsilon U_-,0).
\end{equation}
Note that, in the notation above, $U = \gamma v$, where $v$ is the three-speed and $\gamma = \sqrt{1+U^2}$. Hence, $0\leq U<\infty$. In the fluid's rest frame, the cosmic string speed is $v_+$. 

Equation \eqref{RHCSeq2} thus gives
\begin{equation}\label{RHCSnumberdensity}
    n_- U_-(\alpha-\epsilon)= n_+ U_+ \alpha,
\end{equation}
where $\alpha = \tan \Phi$. Meanwhile, the $\mu =0, 1, 2$ components of \eqref{RHCSeq1} give, respectively,
\begin{subequations}\label{RHCSenergy-momentum}
\begin{align}\label{RHCSenergy-momentum1}
    (\rho_- + p_-)(\alpha-\epsilon)\gamma_- U_- &=(\rho_+ + p_+)\alpha\gamma_+ U_+, 
    \\\label{RHCSenergy-momentum2}
    (\rho_- + p_-)(\alpha-\epsilon)U^2_- + \alpha p_-&=\alpha(\rho_+ + p_+)U^2_+ + \alpha p_+,\\\label{RHCSenergy-momentum3}
    p_- + (\rho_- +p_-)\epsilon (\epsilon -\alpha)U^2_-&= p_+ .
\end{align}
\end{subequations}
The last three equations have four unknowns, $\alpha, U_-, \rho_-$, and $p_-$, and so we need an equation of state behind the shock to solve for them exactly. Having found $U_-$ and $\alpha$, we might solve \eqref{RHCSnumberdensity} for $n_-$. However, for weak shocks, in which the change in the fluid variables across the shock is small, we can relate energy density and pressure perturbations using the fluid's adiabatic sound speed. In that case, we can find the variation in the fluid variables for an arbitrary equation of state. Such solutions should exist, for a fixed $U_+$, provided the deficit angle is small enough and the change in the fluid variables are of order $G \mu$.

To find the weak shock solution, we write
\begin{equation}
    U_- = U_+ + \delta U, \quad \rho_- = \rho_+ +\delta \rho, \quad p_- = p_+ +\delta p,
\end{equation}
and make the approximation
\begin{equation}
    \delta p \approx c_s^2 \delta \rho,
\end{equation}
where
\begin{equation}
    c_s^2 = \left(\frac{\partial p}{\partial \rho}\right)_{s}
\end{equation}
is the fluid's adiabatic sound speed. Now, we need to solve equations \eqref{RHCSenergy-momentum} to first order in $G \mu$. Firstly, we use equations \eqref{RHCSenergy-momentum2} and \eqref{RHCSenergy-momentum3} to write $p_-$ in terms of known quantities and $\alpha$,
\begin{equation}\label{RHCSpminus}
    p_- =p_+ + \frac{\epsilon\alpha}{1+\epsilon \alpha}(\rho_+ + p_+)U^2_+.
\end{equation}
Then we plug this result into \eqref{RHCSenergy-momentum2} to find
\begin{equation}\label{minusvsplus}
    U_-^2(\rho_- +p_-)(\alpha-\epsilon) = \frac{\alpha}{1+\epsilon \alpha}(\rho_+ + p_+)U^2_+,
\end{equation}
and inserting this into \eqref{RHCSenergy-momentum1} yields, after some algebraic manipulations,
\begin{equation}\label{RHCSUminus}
    U_-^2 = \frac{U_+^2}{(1+\epsilon\alpha)^2(1+U_+^2)-(1+\epsilon^2)U_+^2}.
\end{equation}
Perturbing \eqref{RHCSpminus} and \eqref{RHCSUminus} we find, to first order in $\epsilon$,
\begin{align}
\delta U &\approx - \epsilon \alpha U_+(1+U_+^2), \\
\delta p &\approx \epsilon \alpha (\rho_+ + p_+)U_+^2.
\end{align}
Using these results and one of the original equations, we can find $\alpha$ in terms of $c_s$ and known quantities. For instance, perturbing \eqref{RHCSenergy-momentum2} gives
\begin{equation}
    \alpha\left[(\delta \rho + \delta p)U^2_+ + 2(\rho_+ +p_+)U_+ \delta U + \delta p\right] - \epsilon (\rho_+ + p_+)U_+^2 =0,
\end{equation}
to first order in $\epsilon$. Inserting the expressions for $\delta U$ and $\delta p$ into this equation yields
\begin{equation}
    \tan \Phi = \alpha \approx \frac{U_s}{\sqrt{U^2_+ - U_s^2}},
\end{equation}
where $U_s = \gamma_s c_s = c_s/\sqrt{1-c_s^2}$. Note that for $U_+ \sim U_s$, $\Phi \simeq \pi/2$, which matches the high-speed limit of subsonic flow \cite{Hiscock:1988zn}. However, for string speeds much higher than the sound speed, $\alpha$ decreases and the shock becomes a thin wedge in the wake of the string through the fluid.

In summary, to first order in the deficit angle, the linearized shock solution to the relativistic Taub-Rankine-Hugoniot equations is \cite{Hiscock:1988zn}
\begin{align}
\delta U &\approx - \epsilon \alpha U_+(1+U_+^2), \\
\delta p &\approx \epsilon \alpha (\rho_+ + p_+)U_+^2, \\
\delta \rho &\approx \epsilon \alpha (\rho_+ + p_+)(1+U_s^2)\frac{U_+^2}{U_s^2},\\
\alpha &\approx \frac{U_s}{\sqrt{U^2_+ - U_s^2}}.
\end{align}

The linearized solution above cannot be trusted for string speeds very close to the speed of light, or for very non-relativistic sound speeds: in those cases, the perturbations are not so small compared with the values outside the shock. For instance, the condition $\delta \rho < (\rho_+ + p_+)$ gives
\begin{equation}
    \epsilon \frac{U_s}{\sqrt{U_+^2 -U_s^2}}(1+U_s^2)\frac{U_+^2}{U_s^2}<1,
\end{equation}
which can be violated as $U_+/U_s \to \infty$ for a fixed $\epsilon$. In this limit, the shock is very strong and $\alpha$ becomes very small, of the order of $\epsilon$. Let us assume that this is the case and estimate the change in the fluid variables after taking $\alpha$ to be larger but of the same order of the deficit angle:
\begin{equation}
    \alpha \sim 4\pi G \mu, \quad \alpha - 4\pi G \mu \sim 4\pi G \mu.
\end{equation}
So, expanding \eqref{RHCSUminus} for large $U_+$ and $\alpha \sim \epsilon \approx 4\pi G \mu \ll1$, we get
\begin{equation}
    U_-^2 \sim \frac{1}{3\epsilon^2} \sim \frac{1}{48 \pi^2 G^{2} \mu^2},
\end{equation}
while \eqref{RHCSpminus} gives
\begin{equation}
    p_- \sim 2\epsilon^2 (\rho_+ +p_+)U_+^2 \sim 32\pi^2 G^{2} \mu^2(\rho_+ +p_+)U_+^2.
\end{equation}
Using these results into \eqref{minusvsplus}, we also find
\begin{equation}
    \rho_- \sim 4 \epsilon^2 (\rho_+ +p_+)U_+^2 \sim 64 \pi^2 G^{2} \mu^2 (\rho_++p_+)U_+^2.
\end{equation}
Hence, we conclude that, for $U_+ \gg \left(G \mu\right)^{-1}$, the jump in energy density and pressure is very large. Presumably, the jump in temperature is also very big, and the BEC cannot be maintained inside the shock. In fact, at very low temperatures, the equation of state for a gas of weakly interacting bosonic particles has a weak dependence on the number density of particles. In the non-interacting case, $p \propto T^{5/2}$, and so a large increase in pressure is accompanied by a large increase in temperature. More realistically, we should also take into account the change in number density, since it might be large enough to imply a higher value of $T_c$ inside the strong shock. However, from \eqref{RHCSnumberdensity}, we get
\begin{equation}
    n_- = \frac{n_+ U_+}{U_-} \frac{\alpha}{\alpha-\epsilon} \sim 2\sqrt{3}\; n_+ U_+ \epsilon,
\end{equation}
and so, although significant, the fractional change in the number density is comparatively smaller than in the
ones associated with energy density and pressure. Hence, we conclude that, generically, the temperature in the strong shock will increase, such that $T_-> T_c$ and the BEC is destroyed in the wake of the moving string.

Fortunately, for DM condensed in the superfluids phase, the sound speed is non-relativistic and, moreover, the typical speed of large sections of long cosmic strings is not so close to the speed of light. So, realistically, the linearized solution is a good approximation for the fluid variables inside the shock. In fact, the typical sound speed in the core of spherical DM superfluid condensates is $c_s \sim 10^{-5}c$ (as derived from the solution in \cite{Berezhiani:2015bqa}). In this case, the condition $\delta \rho< (\rho_+ + p_+)$ for the linearized solution to be consistent gives
\begin{equation}
    U_+ < \frac{c_s}{4\pi G \mu},
\end{equation}
and so $U_+ < 10^{-6} \left(G \mu\right)^{-1}$. For $G \mu \sim 10^{-7}$, we get $U_+ < 1$, and so weak shocks require a Lorentz factor associated to $v_+$ at most of order unity, which is the case for cosmic string speeds.

\section{Superfluid phase inside cosmic string wakes}\label{superfluidsinwakes}

In this section, we want to analyse under which conditions a BEC can be formed inside the wake of a cosmic string. For the condensate to be formed, two conditions need to be satisfied \cite{LLfluids,Berezhiani:2015bqa}: first, the thermal de Broglie wavelength $\lambda_{\mathrm{th}}$ of the DM particles has to be larger than the characteristic inter-particle separation $l = n^{-1/3}$, and second, the DM particles have to thermalize.

The first condition,
\begin{align}
	\lambda_{\mathrm{th}} = \sqrt{\frac{2 \pi \hbar^{2}}{m k_{B} T}} \gtrsim l = \left(\frac{1}{n}\right)^{1/3},
\end{align}
is equivalent to requiring that the de Broglie wavelength of DM particles overlap in the region where we want them to condense. Equivalently, at a fixed temperature $T$, the BEC forms when the number density of particles is larger than a critical density $n_{c}$\cite{Khoury:2021tvy},
\begin{align}
    n > n_{c} \equiv \frac{\zeta\left(\frac{3}{2}\right)}{\lambda_{\mathrm{th}}^{3}},
\end{align}
where $\zeta\left(3/2\right)$ is the Riemmann zeta function. Assuming that the DM particles follow the velocity dispersion $k_{B} T \sim m v^{2}/2$ and using $\rho = nm$, the previous inequality reduces to the following bound on the mass:
\begin{align}
	m^{4} \lesssim \frac{\left(4 \pi \hbar^{2}\right)^{3/2}}{\zeta\left(3/2\right)} \frac{\rho}{v^{3}}.\label{cond1}
\end{align}

Now, let us apply this condition to the wake of a long string. We consider a wake formed at a time $t_{i} > t_{\mathrm{eq}}$ (wakes formed $\sim t_{\mathrm{eq}}$ have the largest surface density). Due to the passing of the string, the comoving coordinates $x^i$ of the DM particles are perturbed relative to the Hubble flow, such that their physical position is $a\left(t\right) \left[x^i + \psi^i (x^j,t)\right]$. DM accretion into wakes can be described by the time evolution of $\psi^{i}$. Assuming the Zel'dovich approximation and the wake along the $x$-axis, such that only $\psi = \psi^y$ is non-trivial, we have \cite{vilenkin1994cosmic}
\begin{equation}
    \psi(t,y) = -\frac{3}{5}u_i t_i\left[\left(\frac{t}{t_i}\right)^{2/3}- \left(\frac{t}{t_i}\right)^{-1}\right] \frac{\left[\theta(y)- \theta(-y)\right]}{2},
\end{equation}
where $\theta(y)$ is the Heaviside step function and $u_i = a(t_i)^{-1} 4 \pi G \mu v_s \gamma_s$ is an initial velocity boost given by the string to nearby particles. Here, $v_{s}$ is the velocity of the string, and $\gamma_{s} = \left(1-v_{s}^{2}\right)^{-1/2}$ the corresponding Lorentz factor.  For $t\gg t_i$, only the first term in the brackets significantly contributes to the solution.

The physical distance of a DM particle to the wake, which initially increases because of the Hubble flow, becomes maximal at a time $\bar{t}$ and eventually starts decreasing due to the gravitational pull of shells of matter. So, we have DM shells that turn around at $\Bar{t}$. The physical height of the turnaround surface above the wake's center is
\begin{equation}
    h(\bar{t}) = a(\bar{t}) |\psi(\bar{t})| = \frac{3}{5}u_i t_i \left(\frac{\bar{t}}{t_i}\right)^{2/3} \left(\frac{\bar{t}}{t_0}\right)^{2/3}.
\end{equation}

The velocity of the particles inside the virialized region in the wake is 
\begin{align}
	v_{\text{vir}} \equiv v\left(t_{v}\right) =  a|\Dot{\psi}(t_v)| =\frac{2}{5} u_{i} \left(\frac{t_{i}}{t_{0}}\right)^{1/3} \left(\frac{t_{v}}{t_{0}}\right)^{1/3} = \frac{2^{3/4}}{5} u_{i} \left(\frac{t_{i}}{t_{0}}\right)^{1/3} \left(\frac{\bar{t}}{t_{0}}\right)^{1/3},\label{wakevel}
\end{align}
where $\Bar{t}$ is the time when a shell of particles turns around and falls into the wake, and $t_{v} = 2^{-3/4} \bar{t}$ is the time when the shell enters the virialized region, which we assume to have a height of $h(\Bar{t})/2$ above the wake's center. The density of DM particles inside this region,  $\rho_{\text{vir}}$, is four times the background density.

\begin{figure}[t]
	\centering
	\includegraphics[scale=.6]{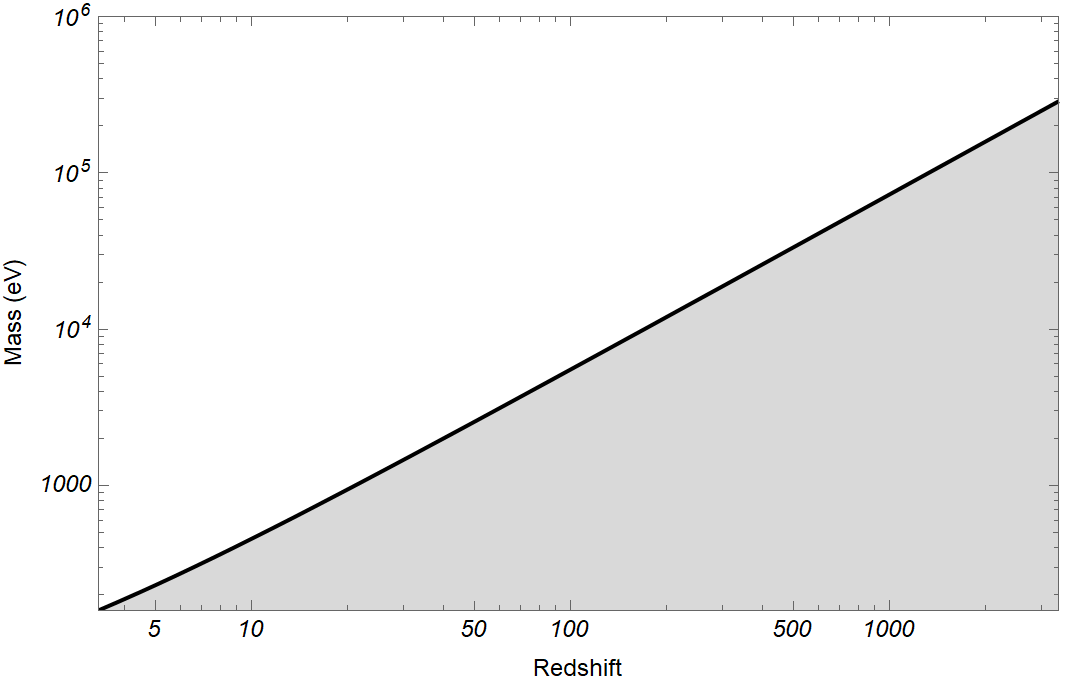}
	\caption{Log-log plot of the upper bound on the DM-particle mass \eqref{massbound} as a function of the redshift, $m\left(z\right)$. The shaded region shows the range of allowed masses for the particles' de Broglie wavelength to be larger than the typical inter-particle separation inside the cosmic string wake.}
	\label{f02}
\end{figure}

Assuming matter domination and $H_{0} = h \times 2 \times 10^{-33} \mathrm{eV}$, we obtain an upper bound on the mass of the DM particles,
\begin{align}\label{massbound}
	m &\lesssim 31 \left(\frac{h}{0.67}\right)^{1/2} \left(\frac{G \mu}{10^{-7}}\right)^{-3/4} \left(\frac{v_{s} \gamma_{s}}{1/\sqrt{3}}\right)^{-3/4} \left(\frac{1 + z_{i}}{1 + z_{\mathrm{eq}}}\right)^{-3/8} \left(1 + \bar{z}\right)^{9/8} \mathrm{eV}.
\end{align}
Figure~\ref{f02} shows a plot of the upper bound on the mass $m$ in eV as a function of the redshift $z$; the shaded region is the range of allowed masses. Hence, at around the epoch of reionization, $z \sim 10$, the DM particles have to be lighter than $\sim 455$~eV for the BEC to form.

The second condition, the requirement that the DM particles thermalize, can be written as
\begin{align}
	\Gamma t_{\mathrm{dyn}} \gtrsim 1,
\end{align}
where $\Gamma$ is the DM self-interaction rate, and $t_{\mathrm{dyn}}$ the time associated to the wake dynamics. For scalar DM, the former is given by \cite{Sikivie:2009qn}
\begin{align}
	\Gamma &= \mathcal{N} v \rho \frac{\sigma}{m},	\qquad	\mathcal{N} = \frac{\rho}{m} \frac{\left(2 \pi\right)^{3}}{\frac{4 \pi}{3} \left(m v\right)^{3}},
\end{align}
where $\sigma$ is the DM self-interaction cross-section and the Bose enhancement factor $\mathcal{N}$ takes into account the fact that the DM particles interact over an excited background state. The dynamical time can be estimated as the time it takes for a DM particle to cross the virialized region,
\begin{equation}
    t_{\text{dyn}} \approx \frac{h(\bar{t})}{v_{\text{vir}}} = \frac{3}{2^{3/4}}\Bar{t} = \frac{3}{2^{3/4}} t_0 (1+\Bar{z})^{-3/2}.
\end{equation}
Combining those expressions, we obtain a bound on $\sigma/m$ of
\begin{align}\label{crosssecbound}
	\frac{\sigma}{m} &\gtrsim 4 \times 10^{-2} \left(\frac{m}{\mathrm{eV}}\right)^{4} \left(\frac{G \mu}{10^{-7}}\right)^{2} \left(\frac{v_{s} \gamma_{s}}{1/\sqrt{3}}\right)^{2} \left(\frac{h}{0.67}\right)^{-3} \left(\frac{1 + z_{i}}{1 + z_{\mathrm{eq}}}\right) \left(1 + \bar{z}\right)^{-11/2} \frac{\mathrm{cm}^{2}}{\mathrm{g}}.
\end{align}
Equivalently, $\sigma/m \gtrsim 2 \times 10^{2}$~GeV$^{-3}$, for the same fiducial value of the parameters.

We can estimate the critical temperature of the condensate by computing the critical velocity $v_c$ that saturates the first condition on the mass,
\begin{equation}
    v_c = \frac{(4\pi \hbar^2)^{1/2}}{\zeta^{1/3}(3/2)}\left(\frac{\rho}{m^4}\right)^{1/3},
\end{equation}
and then use $T_c \sim m v_c^2/(2k_B)$:
\begin{equation}
    T_c \sim \frac{m}{2k_B}\frac{(4\pi \hbar^2)}{\zeta^{2/3}(3/2)} \left(\frac{\rho}{m^{4}}\right)^{2/3} = 11 \left(\frac{m}{\text{eV}}\right)^{-5/3} \left(\frac{h}{0.67}\right)^{4/3}\left(1+\bar{z}\right)^2 \; \text{mK}.
\end{equation}

For $T<T_c$, we have the superfluid phase. We can estimate the fraction of DM particles in the BEC component after neglecting interactions. In this case, the fraction is just the one for an ideal quantum gas of bosonic particles,
\begin{equation}
    \frac{N_0}{N} \approx 1- \left(\frac{T}{T_c}\right)^{3/2}.
\end{equation}
The gas temperature in units of $T_c$ is given by
\begin{equation}
    \frac{T}{T_c} \sim 1 \times 10^{-4} \left(\frac{h}{0.67}\right)^{-4/3} \left(\frac{m}{\text{eV}}\right)^{8/3}\left(\frac{G \mu}{10^{-7}}\right)^2 \left(\frac{v_s\gamma_s}{1/\sqrt{3}}\right)^{2}\left(\frac{1+ z_i}{1+z_{\text{eq}}}\right)(1+\Bar{z})^{-3},
\end{equation}
and so
\begin{equation}
    \frac{N_0}{N} \approx 1- 1 \times 10^{-6} \left(\frac{h}{0.67}\right)^{-2} \left(\frac{m}{\text{eV}}\right)^{4}\left(\frac{G \mu}{10^{-7}}\right)^3 \left(\frac{v_s\gamma_s}{1/\sqrt{3}}\right)^{3}\left(\frac{1+ z_i}{1+z_{\text{eq}}}\right)^{3/2}(1+\Bar{z})^{-9/2}.
\end{equation}
Thus, most of the particles are found in the BEC, regardless of the turnaround redshift.

\section{Discussions and Conclusion}\label{conclusion}

In this paper we studied the motion of a cosmic string through a gas of superfluid dark matter particles. We first reviewed the necessary formalism to approach the subject, an effective field theory approach to superfluids. We then studied two distinct cases: a cosmic string passing through a BEC, and a string moving through a region where the DM is not condensed. In the first case, we looked at the shock induced in the fluid by the string and solved the Taub-Rankine-Hugoniot junction equations. For usual cosmic string speeds, we concluded that the shock is weak and therefore the DM remains in the superfluid phase after the passage of the string. For extreme cases in which the cosmic string travels at velocities very close to $c$, we found that the large jump in energy and pressure across the shock leads to an increase in the temperature and the subsequent destruction of the condensate. 

In the second case, we studied under which conditions the DM might condense into a superfluid phase. A string moving through a fluid leads to the formation of an overdensity in its wake. Similarly to what happens in galaxies, this increase in the DM density can cause it to condense into a superfluid, provided that two conditions are satisfied: first, the de Broglie wavelengths of the DM particles have to overlap inside the wake, and second, the particles have to thermalize. The former condition was translated into an upper bound on the mass of the DM particles, $m \lesssim 31$~eV for a wake formed at $z_{\mathrm{eq}}$, $G\mu \sim 10^{-7}$ and string speeds $\sim 0.5$. The latter condition led to a lower bound on the ratio between the interaction cross-section and mass, $\sigma/m \gtrsim 4 \times 10^{-2} (m/\text{eV})^4\; \text{cm}^2/\text{g}$ for the same parameters, in agreement with constraints on the cross-section of self-interacting DM, $\sigma/m < \text{cm}^2/\text{g}$ \cite{Randall_2008}. As can be seen from \cite{Berezhiani:2015bqa}, these bounds are compatible with the ones in models of superfluid DM in galactic scales. For sub-eV particles, taking into account the Bose-enhancement factor in the interaction rate, the Bullet Cluster constraint gives $\sigma/m \lesssim 10^{-2}(m/\text{eV})^4 \;\text{cm}^2/\text{g}$ \cite{Berezhiani:2021rjs, Berezhiani:2022buv}\footnote{We thank Lasha Berezhiani for bringing this point to our attention.}. From \eqref{crosssecbound}, we see that there are regions of the $(G\mu, v_s, z_i,\bar{z})$ parameter space that can easily accommodate this refined constraint. Finally, we computed the critical temperature below which de DM condenses, and the result is in the mK range for the same parameters used to estimate the previous bounds.

As future directions, one could study how the presence of a BEC inside the cosmic string wake leads to new observational signatures. A new baryonic interaction that arises from the coupling to the phonons would modify the equations describing baryonic accretion. This should lead to changes in the thickness and/or shape of the wake, and in consequence, to modifications in the wake signatures. Such new features would affect wake signals in 21-cm surveys, CMB polarization, and large-scale structure maps \cite{Danos:2010gx,Brandenberger:2010hn,Hernandez:2012qs,NevesdaCunha:2018urr,Laliberte:2018ina,Maibach:2021qqf,Blamart:2022kly}. As mentioned, cosmic strings moving through a condensate at ultra-relativistic speeds are expected to destroy the condensate. Since the typical speeds of long cosmic strings are much lower, a more relevant scenario might be oscillating loops in a condensate: in the context of halo accretion, loops oscillating at speeds close to $c$ will heat up the fluid above $T_{c}$ and destroy the condensate. This affects the  usual scenario of DM accretion into loops. 

On a more speculative note, the general dynamics of the superfluid around moving cosmic strings might be such that vortices are formed. The general motion of vortices in the conical geometry transverse to cosmic strings has been studied in \cite{Gibbons1989}, although in the non-relativistic limit. Since cosmic strings are relativistic, it would be interesting to generalize these results. Superfluid vortices generated by cosmic string motion would contribute to the spin of cosmic filaments, as studied in \cite{Alexander:2021zhx}, and leave strong-lensing observable imprints on the dark matter halo substructure \cite{Alexander:2019puy}. Moreover, it would be worth exploring how a change in the nature of the superfluid could modify our results, such as in the dark-charged superfluid model of \cite{Alexander:2018fjp, Alexander:2020wpm}.

\section*{Acknowledgments}

We would like to thank Stephon Alexander for comments on an early version of this work. H.B. was supported by the Fonds de recherche du Qu\'ebec (PBEEE/303549). Research at McGill is partially supported by funds from NSERC and the Canada Research Chair program.





\bibliographystyle{bibstyle} 
\bibliography{references.bib}






\end{document}